\documentclass[prb,twocolumn,showpacs,preprintnumbers,amsmath,aps]{revtex4}
\usepackage{graphicx}
\usepackage{dcolumn}
\usepackage{bm}
\usepackage{subfigure}

\newcommand\vp{\varphi}

\newcommand\des{{\hbox{\begin{scriptsize}d\end{scriptsize}}}}
\newcommand\pure{{\hbox{\begin{scriptsize}p\end{scriptsize}}}}
\newcommand\stable{{\hbox{\begin{scriptsize}s\end{scriptsize}}}}
\newcommand\harris{{\hbox{\begin{scriptsize}H\end{scriptsize}}}}

\newcommand\vpu{\vec\varphi_1}
\newcommand\vpd{\vec\varphi_2}
\newcommand\tr{\text{Tr}}
\def\nbC{{\mathchoice {\setbox0=\hbox{$\displaystyle\rm C$}%
\hbox{\hbox to0pt{\kern0.4\wd0\vrule height0.9\ht0\hss}\box0}}
{\setbox0=\hbox{$\textstyle\rm C$}\hbox{\hbox
to0pt{\kern0.4\wd0\vrule height0.9\ht0\hss}\box0}}
{\setbox0=\hbox{$\scriptstyle\rm C$}\hbox{\hbox
to0pt{\kern0.4\wd0\vrule height0.9\ht0\hss}\box0}}
{\setbox0=\hbox{$\scriptscriptstyle\rm C$}\hbox{\hbox
to0pt{\kern0.4\wd0\vrule height0.9\ht0\hss}\box0}}}}
\def\nbQ{{\mathchoice {\setbox0=\hbox{$\displaystyle\rm
Q$}\hbox{\raise
0.15\ht0\hbox to0pt{\kern0.4\wd0\vrule height0.8\ht0\hss}\box0}}
{\setbox0=\hbox{$\textstyle\rm Q$}\hbox{\raise
0.15\ht0\hbox to0pt{\kern0.4\wd0\vrule height0.8\ht0\hss}\box0}}
{\setbox0=\hbox{$\scriptstyle\rm Q$}\hbox{\raise
0.15\ht0\hbox to0pt{\kern0.4\wd0\vrule height0.7\ht0\hss}\box0}}
{\setbox0=\hbox{$\scriptscriptstyle\rm Q$}\hbox{\raise
0.15\ht0\hbox to0pt{\kern0.4\wd0\vrule height0.7\ht0\hss}\box0}}}}
\def\nbT{{\mathchoice {\setbox0=\hbox{$\displaystyle\rm
T$}\hbox{\hbox to0pt{\kern0.3\wd0\vrule height0.9\ht0\hss}\box0}}
{\setbox0=\hbox{$\textstyle\rm T$}\hbox{\hbox
to0pt{\kern0.3\wd0\vrule height0.9\ht0\hss}\box0}}
{\setbox0=\hbox{$\scriptstyle\rm T$}\hbox{\hbox
to0pt{\kern0.3\wd0\vrule height0.9\ht0\hss}\box0}}
{\setbox0=\hbox{$\scriptscriptstyle\rm T$}\hbox{\hbox
to0pt{\kern0.3\wd0\vrule height0.9\ht0\hss}\box0}}}}
\def\nbS{{\mathchoice
{\setbox0=\hbox{$\displaystyle     \rm S$}\hbox{\raise0.5\ht0%
\hbox to0pt{\kern0.35\wd0\vrule height0.45\ht0\hss}\hbox
to0pt{\kern0.55\wd0\vrule height0.5\ht0\hss}\box0}}
{\setbox0=\hbox{$\textstyle        \rm S$}\hbox{\raise0.5\ht0%
\hbox to0pt{\kern0.35\wd0\vrule height0.45\ht0\hss}\hbox
to0pt{\kern0.55\wd0\vrule height0.5\ht0\hss}\box0}}
{\setbox0=\hbox{$\scriptstyle      \rm S$}\hbox{\raise0.5\ht0%
\hboxto0pt{\kern0.35\wd0\vrule height0.45\ht0\hss}\raise0.05\ht0%
\hbox to0pt{\kern0.5\wd0\vrule height0.45\ht0\hss}\box0}}
{\setbox0=\hbox{$\scriptscriptstyle\rm S$}\hbox{\raise0.5\ht0%
\hboxto0pt{\kern0.4\wd0\vrule height0.45\ht0\hss}\raise0.05\ht0%
\hbox to0pt{\kern0.55\wd0\vrule height0.45\ht0\hss}\box0}}}}
\def\nbZ{{\mathchoice {\hbox{$\sf\textstyle Z\kern-0.4em Z$}}
{\hbox{$\sf\textstyle Z\kern-0.4em Z$}}
{\hbox{$\sf\scriptstyle Z\kern-0.3em Z$}}
{\hbox{$\sf\scriptscriptstyle Z\kern-0.2em Z$}}}}

\begin{document}

\title{Competition between fluctuations and disorder in frustrated magnets}

\author{Julien Serreau}
\email{serreau@thphys.uni-heidelberg.de}
\affiliation{Institut f\"ur Theoretische Physik, Universit\"at
 Heidelberg, Philosophenweg 16, 69120 Heidelberg, Germany}

\author{Matthieu Tissier}
\email{tissier@thphys.uni-heidelberg.de}
\affiliation{Institut f\"ur Theoretische Physik, Universit\"at
 Heidelberg, Philosophenweg 16, 69120 Heidelberg, Germany}

\date{\today}

\begin{abstract}
We investigate the effects of impurities on the nature of the phase
transition in frustrated magnets, in $d=4-\epsilon$ dimensions. For
sufficiently small values of the number of spin components, we find no
physically relevant stable fixed point in the deep perturbative region
($\epsilon \ll 1$), contrarily to what is to be expected on very
general grounds. This signals the onset of important physical effects.
\end{abstract}


\maketitle

The influence of disorder in solid state physics is a very important
problem. In many cases, it is expected that inhomogeneities, such as
impurities, defects in the lattice structure, etc. may induce a
completely different behavior in the system. Let us discuss two such
situations in the context of phase transitions, which we shall be
concerned with in this report. In systems undergoing second order
phase transitions, disorder most often changes a divergence in the
specific heat at the critical temperature into a cusp. The archetype
of such a situation is the $d=3$ Ising model, for which both
experimental and theoretical works show that the critical exponent
$\alpha$ changes from positive to negative when disorder is added (for
a review, see Ref.\ \onlinecite{folk01b}). This is well understood, thanks to the
Harris criterion,\cite{harris74} which states that for a large class
of disordered systems,\cite{chayes86} the critical exponent
$\nu_\des$ -- which governs the singularity of the correlation length
-- must satisfy the inequality $\nu_\des\geq 2/d$, where $d$ is the
number of spatial dimensions. Using the hyperscaling relation
$\alpha=2-\nu d$, this inequality translates into $\alpha\leq 0$,
which in turn implies that there is no divergence in the specific
heat. More generally, Harris criterion leads to the conclusion that
adding disorder to a pure system with $\nu_\pure\leq 2/d$ has a
dramatic effect, since it must change the critical exponents, and
therefore the universality class of the phase transition.

The influence of disorder on first-order phase transitions has also
been considered for a long time. It was soon proposed that impurities
tend to reduce the discontinuities of the first derivatives of the
free energy (jump of the magnetization, latent heat). It was argued
that these discontinuities could even disappear in some
cases.\cite{imry79} Again disorder changes qualitatively the
properties of the system, by inducing a second-order phase
transition. Later on, this ``rounding effect'' of disorder was proven
to take place in a large class of systems by Aizenman and
Wehr\cite{aizenman89} (referred to as AW in the following). Up  to
now there is no experimental realization of this phenomenon.

It was recently proposed by one of us\cite{tissier02b} that such a
second-order phase transition induced by impurities could actually be
studied by experimental means in frustrated magnets such as stacked
triangular antiferromagnets (CsMnBr$_3$, CsMnI$_3$, etc.) and rare-earth 
helimagnets (Ho, Dy, Tb). In the pure case, these materials have
been very much studied both experimentally and theoretically since the
1970s (see Refs.\ \onlinecite{collins97} and \onlinecite{kawamura98} for 
reviews). The early calculations indicate a second-order phase transition 
(see also Ref.\ \onlinecite{pelissetto01}), in accordance with the observed 
power-law behavior of thermodynamic quantities. However, there is a wide
dispersion of the measured critical exponents. This lack of
universality has been recently interpreted as a signature of a weakly
first-order phase transition.\cite{itakura01,tissier01} This issue is
still very debated. In any case, we believe that it would be very
instructive to study the influence of disorder in such systems. On the
one hand, if the transition is of second order in the pure case, one
expects a dramatic change of the critical exponents when impurities
are added. Following Harris criterion, the exponent $\nu$, which is
always found to be smaller than $0.57$, should jump above $2/3$. This
variation is at least three times bigger than in the case of Ising
magnets and might be easier to observe experimentally.  On the other
hand, if the transition is weakly of first order, it is expected to be
turned into a continuous one in the presence of impurities, on the
basis of AW. As a consequence, one would observe universal critical
exponents, in contrast with the scattered values reported in the pure
case.  This would provide the first experimental realization of the
rounding effect of disorder. This last scenario is supported by a
recent nonperturbative renormalization-group calculation in
$d=3$.\cite{tissier02b} Actually, the results of AW are not restricted to 
$d=3$, and apply to any dimension lower than four.\footnote{Strictly speaking,
this holds if the number of spin components is larger than two} In
the present report, we study the rounding effect of disorder within
the standard $\epsilon$ expansion. As we shall see, the perturbative
results do not coincide with the conclusions of AW, signaling the
onset of physical effects, yet to be determined.

In the following, we describe the one-loop result, from which we argue
that higher-order contributions are needed in order to get the
qualitative behavior of the system. We show that the
$\epsilon$ expansion of the fixed-point coupling constants is
singular for some physically relevant values of the number of spin
components. Using a simple geometrical picture, we explain the origin
of such singularities, and recast the perturbative series into a
regular form. Our discussion is quite general, and can be applied to a
wide class of theories displaying the same qualitative behavior. We
then perform a two-loop calculation and analyze the effect of disorder
using this method.

Let us first describe the Hamiltonian which is used in the
following. The theoretical studies of the phase transition in the
materials considered here started in the 1970s.\cite{garel76,yosefin85} 
It was then realized that the relevant order
parameter is not a simple vector (as for a ferromagnetic-paramagnetic
phase transition), but a set of two vectors $\vpu$ and $\vpd$. This is
a direct consequence of frustration, which makes the ground state more
degenerate than in the corresponding nonfrustrated system. In
physically relevant situations, $\vpu$ and $\vpd$ are taken to have
two or three components, depending on the anisotropies of the
crystal. Here we consider a direct generalization where the order
parameter is composed of $m$ vectors, each having $n$ spin components,
and we gather these into a $n\times m$ matrix:
$\phi=(\vec{\vp}_1,\cdots,\vec{\vp}_m)$. Considering only the relevant
``$\vp^4$-like'' terms, the hamiltonian reads:
\begin{equation}
\label{action_pur}
\mathcal S(\phi)=\int d^dx \Big[ \frac Z2\tr
\left(\partial ^{\:t}\!\phi.\partial\phi\right)+r
\;\frac{\rho}2+\frac {u_1}8 \rho^2 +\frac{u_2}4\tau \Big]
\end{equation}
where $\rho=\tr (\,^t\!\phi\phi\,)$ and $\tau=
\tr(\,^t\!\phi\phi \, ^t\!\!\phi\phi\,)$. Adding
impurities induces a spatial dependence of the coupling constants,
through the local density of impurities. Using the replica trick, it
is, however, possible to average over disorder, and to study the
properties of the system by considering a translational invariant
Hamiltonian. The price to pay is that we now have to deal with $o$
replicas of the field, and to take the limit $o\rightarrow0$ at the
end of the calculation. The $\vp^4$-like hamiltonian then
reads\cite{tissier02b}
\begin{equation}
\label{action_repliques}
\begin{split}
\widetilde{ \mathcal S}(\{\phi_k\})=\sum_{l=1}^o  \mathcal S(\phi_l) +
\frac{u_3}8 \int &d^dx\Big[\sum_{l=1}^o \rho_l\Big]^2,
\end{split}
\end{equation}
where $l$ indices the different copies of the field. Note that only
the region $u_1\geq0$, $m\,u_1+2u_2\geq0$, $u_3\leq0$ is of physical
relevance.\cite{folk01b,yosefin85}

Let us describe the situation at the one-loop level. The
$\beta$ functions were derived in Ref.\ \onlinecite{tissier02b}, and can be read
off on Eq.~(\ref{beta_pert}) below. Out of the eight solutions to the
set of three equations $\beta_i=0$, only four are of interest
here. The situation of pure frustrated magnets is recovered when
$u_3=0$. In this plane, we retrieve the two associated fixed points,
noted $P^+$ and $P^-$ in the following. In addition to these
well-known fixed points, we find two new ones which are of particular
interest in our study, and which we call $D^+$ and $D^-$. Let us
discuss the behavior of the fixed point when $n$ is varied, begining
with $P^+$ and $P^-$.\cite{garel76} For large values of $n$, they
have real coordinates in coupling-constant space. When $n$ is
decreased, they approach each other, and coincide at the critical
value $n^+_{\pure,0}=4(3+\sqrt 6)\simeq 21.8$ (throughout this report,
the numerical values are given for the physically interesting case
$m=2$). Below $n^+_{\pure,0}$, the two fixed points get complex
coordinates and are not physical anymore. Finally, when $n$ reaches
$n^-_{\pure,0}=4(3-\sqrt 6)\simeq 2.2$, $P^+$ and $P^-$ reappear,
however, not in a region of physical interest. These features follow
directly from the expression of the coupling constants at $P^\pm$,
which read ($i=1,2,3$)
\begin{equation}
u^\pm_i=\epsilon\,A_i(n)\pm \epsilon\,B_i(n)\sqrt{n-n
^-_{\pure,0}}\sqrt{n-n_{\pure,0}^+}.
\label{constantes_une_boucle}
\end{equation}
We observe the very same qualitative behavior for the couple $D^\pm$,
and we denote by $n_{\des,0}^\pm$ the associated critical values of
$n$. Surprisingly, the two couples of fixed points $P^\pm$ and $D^\pm$
annihilate for the {\em same} values of $n$ (namely, $n_{\pure,0}^\pm =
n_{\des,0}^\pm\equiv n_0^\pm$), and at the {\em same} location in
coupling-constant space. In the following, we focus our attention on
the vicinity of the upper critical value $n_0^+$, the only physically
interesting one, and we drop the superscript.

In summary, at one loop, when $n>n_0$, the fixed point $P^+$ is
stable, and the disorder is irrelevant. For $n<n_0$, there is no
stable fixed point, and the phase transition is likely to be of first
order. This is very astonishing, being difficult to reconcile with the
general results of AW. Thus one can expect that higher-order
corrections will resolve this mismatch. In fact, we expect a
qualitative change already at two loops. Indeed, for the pure
frustrated magnets, at the value of $n$ where the two fixed points
$P^\pm$ meet, one has:\cite{pelissetto01b} $\nu=2/d- \epsilon^2(\sqrt
6-1)/200+o(\epsilon^3)<2/d$. Using Harris criterion, we conclude that
the disorder should be relevant, at least for this value of $n$. This
in turn suggests that the degeneracy observed at one loop should be
resolved at higher orders, namely, that the two couples of fixed points
$P^\pm$ and $D^\pm$ should annihilate independently and for different
values of $n$.

\begin{figure}[htbp] 
\centering \makebox[\linewidth]{\hfill
 \subfigure[$n>n_c$]{
\label{poisson_1}
\includegraphics[width=.25\linewidth,origin=tl]{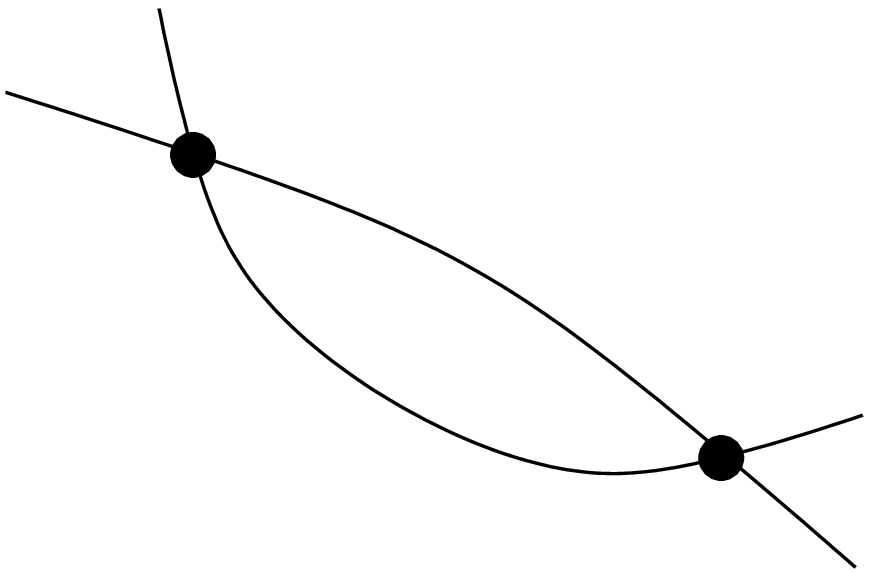}}
\hfill
\subfigure[$n=n_c$]{
\label{poisson_2}
\includegraphics[width=.25\linewidth,origin=tl]{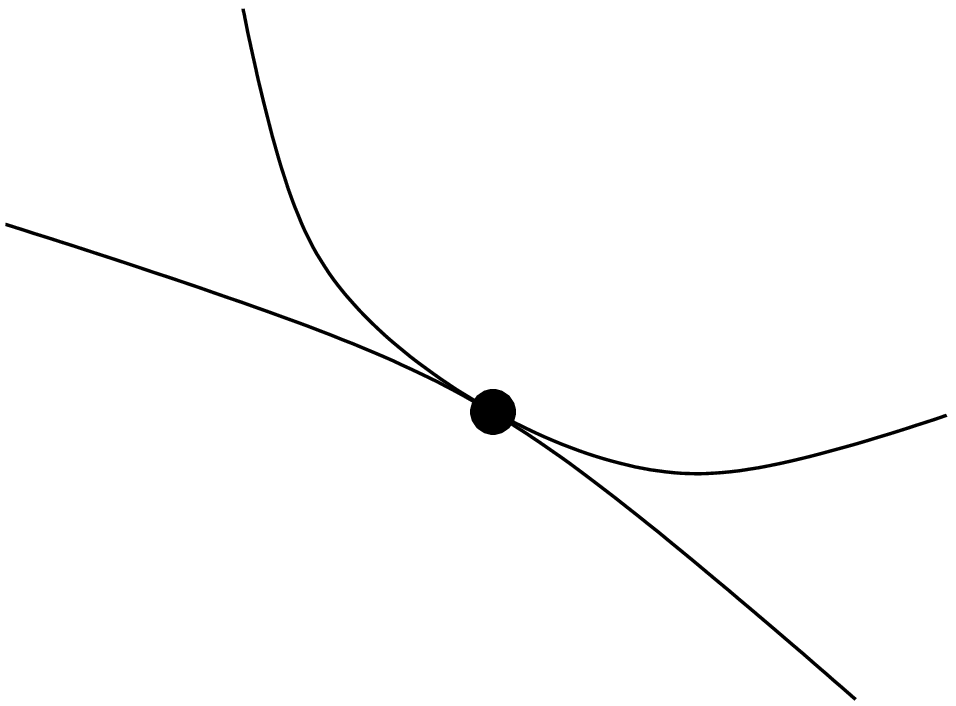}}
\hfill
\subfigure[$n<n_c$]{
\label{poisson_3}
\includegraphics[width=.25\linewidth,origin=tl]{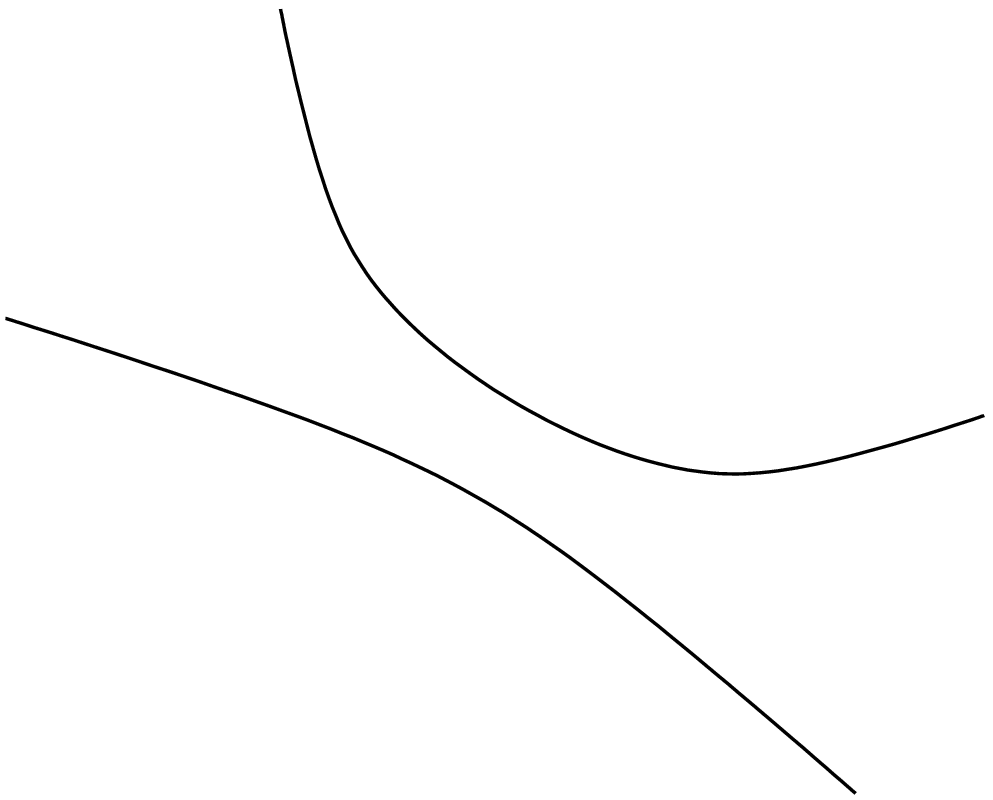}
}\hfill
 }
\caption{Annihilation of two fixed points at $n=n_c$. The lines
represent zeros of the $\beta$ functions and their intersections
represent the fixed points.}
\label{poisson}
\end{figure}

Before embarking in higher-loop calculations, it is useful to do a
short digression and to consider the phenomenon of annihilation of
fixed points on more general grounds. For the sake of simplicity,
consider a theory with two coupling constants $u_1$ and $u_2$. The
points where the associated $\beta$ functions vanish form two curves
in the plane $(u_1,u_2)$, whose intersections are the fixed
points. When a parameter is varied -- in our case the number of spin
components -- these curves move, and so do their intersections. A
typical situation is represented in Fig.\ \ref{poisson}, where one sees
the annihilation of two fixed points for $n=n_c$.
This qualitative picture is already observed at one loop and it is
clear that it does not change when higher-order corrections are
included: the curves of Fig.\ \ref{poisson} are only slightly
displaced. As a consequence, the position $(u_1^\star,u_2^\star)$
where the two fixed-points meet, as well as the corresponding value
$n_c$ of the number of spin components, receive small corrections at
each order of perturbation theory and therefore depend on
$\epsilon$. From these considerations, it is an easy matter to deduce the
generic form of the fixed-point coordinates near $n_c(\epsilon)$.
Expanding the $\beta$ functions in coupling constants up to second
order around $u_i^\star$, one is left with quadratic expressions as in
the one-loop case. The coordinates of the fixed points then read
\begin{equation}
 u_i^\pm = \epsilon \, R_i (n,\epsilon) \pm
 \epsilon \, S_i (n,\epsilon) \, \sqrt{n-n_c(\epsilon)} \, ,
\label{higherloop}
\end{equation}
where $R_i$ and $S_i$ are polynomials in $\epsilon$. The previous
formula is very appealing when compared with the one-loop
form~(\ref{constantes_une_boucle}) in the vicinity of $n_0^+$, the
main difference\footnote{We mention that in principle one can repeat
this construction near $n=n_0^-$, so as to get a form with two square
roots, similar to~(\ref{constantes_une_boucle}).} being the
$\epsilon$ dependence of the critical value of $n$ . The whole
previous argument is very general, and only makes use of the
geometrical picture of Fig.\ \ref{poisson}. In particular, it does not
rely on perturbation theory and Eq.\ (\ref{higherloop}) is, in this
sense, a nonperturbative result. In the following, we make extensive
use of Eq.\ (\ref{higherloop}) which, when combined with perturbation
theory, provides a very powerful tool.

Let us finally make contact with the usual $\epsilon$ expansion of the
fixed-point coupling constants, the coefficients of which are found to
be singular in the limit $n\rightarrow n_0$.\cite{pelissetto01b} We
can easily understand the origin of such singularities in the light of
our previous discussion: writing $n_c (\epsilon) = n_{c,0} + \epsilon
\, n_{c,1} + \cdots$ and expanding Eq.\ (\ref{higherloop}) in powers of
$\epsilon$, one gets terms of the form $\epsilon^\ell
/(n-n_0)^{m-1/2}$, with $m\leq \ell-1$. Therefore, the regular form
(\ref{higherloop}) corresponds to a resummation of infinitely many
singular terms. In practice, one reconstructs the polynomials $R_i$ and
$S_i$ by matching the $\epsilon$ expansion of Eq.\ (\ref{higherloop}) for
$n\neq n_0$ with the corresponding perturbative series.

Comming back to the problem at hand, we now discuss the expected
influence of higher-loop corrections on the one-loop flow diagram. The
critical values of $n$ introduced before, now depend on $\epsilon$,
and we shall refer to them as $n_\pure(\epsilon)$ and
\begin{figure}[htbp] 
\centering 
\makebox[\linewidth]{\hfill
\includegraphics[width=.40\linewidth,origin=tl]{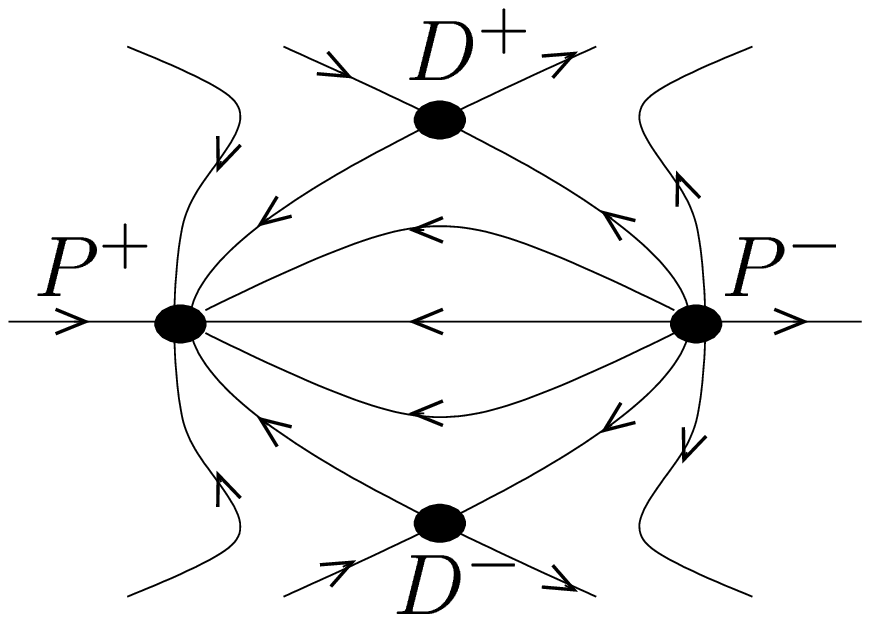} \hfill
\includegraphics[width=.40\linewidth,origin=tl]{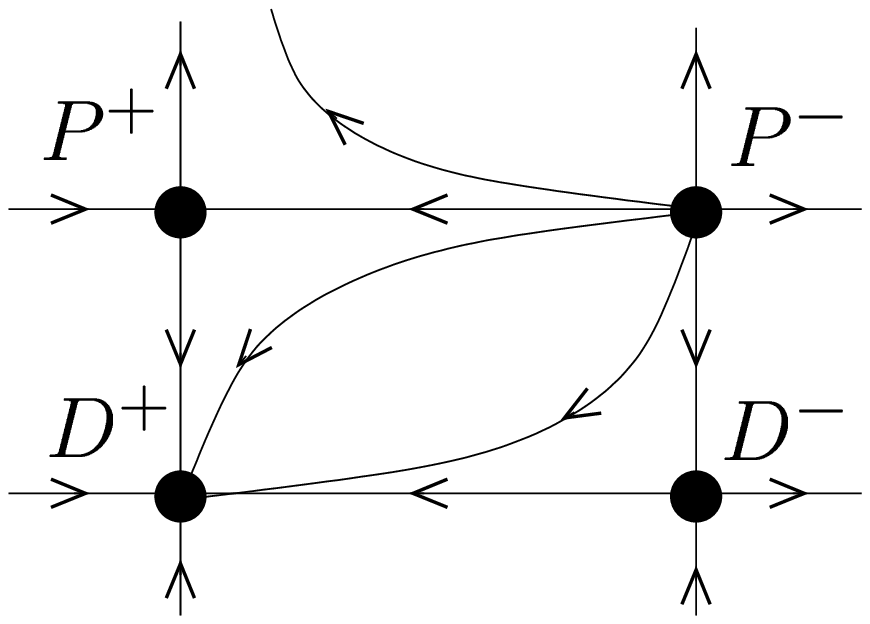}\hfill }
\caption{Flow diagram above and below $n_\text{s}$. The fixed point
$D^+$ crosses the plane $u_3=0$ through $P^+$, and gets stable
afterward. } 
\label{flot_point_fixe} 
\end{figure} 
$n_\des(\epsilon)$. As already explained, Harris criterion indicates
that disorder should be relevant, at least for
$n=n_\pure(\epsilon)$. This can only happen if, for some particular
value $n_\stable(\epsilon)$, either $D^+$ or $D^-$ crosses the plane
$u_3=0$ through $P^+$, and exchanges its stability with it. Moreover,
following AW, we expect that for $n<n_p(\epsilon)$ the phase
transition is governed by a disordered fixed point. This favors the
first scenario, which is depicted in Fig.\ \ref{flot_point_fixe}. In
addition to $n_\pure(\epsilon)$, $n_\des(\epsilon)$, and
$n_\stable(\epsilon)$, we shall be interested in a fourth particular
value of $n$, noted $n_\harris(\epsilon)$, for which the critical
exponent $\nu$ associated with $P^+$ just equals $2/d$. Exploiting the
Harris criterion and the topology of the flow,
we see that $n_\stable$ is larger than
$n_\pure$, $n_\des$, and~$n_\harris$.

With this in mind, we perform a two-loop calculation of the
$\beta$ functions. Using dimensional regularization and minimal
subtraction, we get
\begin{subequations}
\begin{align}
&\begin{aligned} \beta_1&= - u_1\epsilon + 2v_4\big\{
{u_1}^2I_{8,0,1} + 4u_1( 3u_3 + u_2I_{1,1,0} )\\ +&12{u_2}^2 \big\}
- 4{v_4}^2 \big\{3{u_1}^3I_{14,0,3} + 2{u_1}^2( 22u_2I_{1,1,0}\\
+&u_3I_{58,0,11} ) + u_1( 68u_2u_3I_{1,1,0} + 2{u_2}^2I_{87,5,5}\\
+&{u_3}^2J_{82,5} )+ 24{u_2}^2 ( 6u_3 + u_2I_{3,1,0} ) \big\},
\end{aligned}\\ &\begin{aligned} \beta_2&=- u_2\epsilon + 4u_2v_4
\big\{( 6u_1 + u_2I_{4,1,0}+ 6u_3 \big\}\\ -& 4u_2{v_4}^2 \big\{
6{u_2}^2I_{17,3,1} + 4u_2( u_1 + u_3 ) I_{29,11,0}\\ +&u_1 ( u_1 +
2u_3) I_{82,0,5} + {u_3}^2J_{82,5} \big\}, \end{aligned}\\
&\begin{aligned} \beta_3&= - u_3\epsilon + 2u_3v_4 \big\{
2u_1I_{2,0,1} +4u_2I_{1,1,0}\\+& u_3J_{8,1} \big\} - 4u_3{v_4}^2
\big\{ 10{u_2}^2I_{3,1,1}+ 3{u_3}^2J_{14,3}\\+& 4u_2( 5u_1 + 11u_3 )
I_{1,1,0} + u_1( 5u_1 + 22u_3 ) I_{2,0,1} \big\}, \end{aligned}
\end{align}%
\label{beta_pert}%
\end{subequations}
where $v_4=1/(32 \pi^2)$, $I_{i,j,k}=i+j(m+n)+k(m n)$, and
$J_{i,j}=i+j(m n o)$.  We also give the expressions of the critical
exponents:
\begin{subequations}
\begin{align}
&\begin{aligned}
\eta&=2v_4^2\big\{I_{2,0,1}u_1(u_1+2u_3)+ 4I_{1,1,0}u_2(u_1+u_3)\\
&+2I_{3,1,1}u_2^2+2J_{2,1}u_3^2\big\}  \, ,
\end{aligned}\\
&1/\nu=2-2v_4\big\{I_{2,0,1}u_1+2I_{1,1,0}u_2+J_{2,1}u_3\big\}+5\eta
\end{align}
\label{exp_pert}%
\end{subequations}
By using Eq.\ (\ref{higherloop}), we can safely study our expressions near
$n=n_0$. The first surprising result is that the four values of $n$
introduced above are still degenerate at order $\epsilon$, i.e.
$n_{c,1}\equiv n_1= -2(18+7\sqrt{6})/3\simeq-23.4$, for
$c=\text{d, p, s, H}$. In principle, the next correction, i.e. $n_{c,2}$,
is obtained by a three-loop calculation. However, due to 
Eq.\ (\ref{higherloop}) we can show that the degeneracy is resolved at
order $\epsilon^2$ without performing explicitly a three-loop
calculation. Let us illustrate this point on a simple case. When
$n=n_\stable(\epsilon)$, the fixed points $P^+$ and $D^+$ coincide and
the coupling constant $u_3$ of $D^+$ vanishes. Inserting
$n=n_\stable(\epsilon)= n_0 +\epsilon\, n_1 + \epsilon^2
n_{\stable,2}$ in Eq.\ (\ref{higherloop}), and expanding the equation
$u_3^+=0$ to second order in $\epsilon$, one obtains
\begin{equation}
\sqrt{n_{\stable,2}-n_{\des,2}} =- \frac{(\partial_\epsilon+n_1
\partial_n) R_3(n,\epsilon)|_{n=
n_0,\epsilon=0}}{S_3(n_0,0)} \, .
\label{eq_racine}
\end{equation}
The expressions appearing in the above equation involve only
two-loop expressions and can therefore be computed here. For $m=2$, we
find
\begin{equation}
n_{\stable,2}-n_{\des,2}=(1092+463\sqrt 6)/{50}\simeq44.5 \, .
\label{eq_dn2_d}
\end{equation}
Using the same technique for the equations $u_i (D^+)=u_i (P^+)$ at
$n=n_\stable(\epsilon)$ ($i=1,2$), as well as $\nu(P^+)=2/d$ at
$n=n_\harris(\epsilon)$, we get
\begin{equation}
n_{\stable,2}-n_{\pure,2}=n_{\harris,2}-n_{\pure,2}
=(7\sqrt 6-12)/{50}\simeq0.103 \, .
\label{eq_dn2_ph}
\end{equation}
We see that $n_\stable (\epsilon)=n_\harris (\epsilon)$ up to second
order: the Harris criterion is satisfied precisely at this value of
$n$ where disorder becomes relevant. We mention that a three-loop
calculation for the pure ($u_3=0$) frustrated
magnets\cite{pelissetto01b} yielded the value: $n_{\pure,2}\simeq
7.09$.  Combining this with our ``two-loop'' results (\ref{eq_dn2_d})
and (\ref{eq_dn2_ph}), we are able to compute the ``three-loop''
contributions: $n_{\des,2}=-37.3$, $n_{\stable,2}=n_{\harris,2}=~7.19$.
We find the same features for any value of $m$, except for the
numerical results. For example, for $m=3$, we get $n_{\des,2}=-48.2$
and $n_{\stable,2}=n_{\harris,2}=11.1$.

\begin{figure}[htbp] 
\centering 
\includegraphics[width=.75\linewidth,origin=tl]{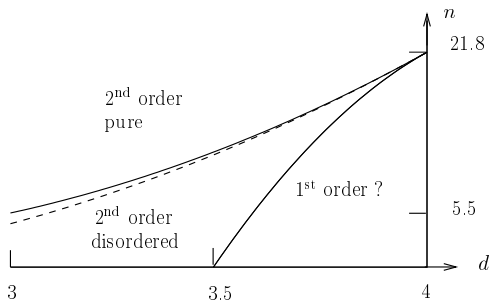}
\caption{Different types of phase transition in the $(n,d)$ plane. The
lines represent $n_\text{\scriptsize s}(\epsilon)$,
$n_\text{\scriptsize p}(\epsilon)$ and $n_\text{\scriptsize
d}(\epsilon)$ from top to bottom. The splitting between
$n_\text{\scriptsize s}(\epsilon)$, and $n_\text{\scriptsize
p}(\epsilon)$ has been magnified}
\label{phases}
\end{figure}

We summarize our results in Fig.\ \ref{phases}.  
There are essentially three distinct regions in the $(n,d)$ plane. In
the first one $[\,n > n_\stable(\epsilon)\,]$, the transition is of second
order and is governed by the pure fixed point $P^+$. In the second one
$[\,n_\des (\epsilon)<n< n_\stable(\epsilon)\,]$, the disorder is relevant
and the system undergoes a second-order phase transition governed by
the fixed point $D^+$. This region is divided by the curve
$n_\pure(\epsilon)$ (dotted line in Fig.\ \ref{phases}). Above it,
$P^+$ still exists, but is unstable against disorder, as expected from
the Harris criterion. Below, the pure fixed point has disappeared: the
fluctuation-driven first-order phase transition observed in the pure
system is turned into a continuous one by impurities, in line with
AW. In the whole second region, the disorder induces a new
universality class, and therefore new critical exponents. For
instance, on the line $n_\des(\epsilon)$, one has $\eta=
\epsilon^2/48$ and $\nu=2/d+\epsilon^2(13+7\sqrt6)/200$ up to
$o(\epsilon^3)$. We note that the experimentally relevant cases
($d=3$, $n=2,3$) belong to this region. Although our results can only
serve as a rough indication in $d=3$, they are corroborated by the
conclusions of a nonperturbative approach, and support the recent
proposal to study experimentally the rounding effect of disorder in
frustrated magnets.\cite{tissier02b} Finally, there is a third
region $[\,n<n_\des(\epsilon)\,]$, where we find no physically relevant
stable fixed point: here, the disorder does not counterbalance the
effect of fluctuations, and the phase transition most probably remains
of first order. We stress that for $n<n_0$, the deep perturbative
regime $\epsilon\ll1$, where our approach is under control, belongs to
this third region.

If the results found in the first two regions are well understood in
terms of the Harris criterion and AW, the situation is still puzzling
in the third region. Indeed, perturbation theory predicts a
discontinuous phase transition where AW state that the latent heat
vanishes, which in turn strongly indicates a continuous transition. We
suggest three possible scenarios which could explain this mismatch. It
could happen that, although the latent heat vanishes, the phase
transition is still of first order\footnote{There could be a jump in
the magnetization, as observed in Ref.\ \onlinecite{gross85}.}. However, 
simple Landau-type arguments indicate that this is not very
likely. Another possibility is that the replica symmetry, which was
implicitly assumed to hold, may actually be broken. This could be
investigated following the lines of Ref.\ \onlinecite{brezin01}. Finally, the
expected second-order phase transition could be governed by a
nonperturbative fixed point, unreachable by the approach we
considered here. These three possibilities are very interesting and
deserve more investigations.

We thank B. Delamotte and D. Mouhanna for
discussions. M. T. acknowledges financial support from Marie Curie
Fellowship Association under contract No. HPMF-CT-2001-01343.


\begin{thebibliography}{16}
\expandafter\ifx\csname natexlab\endcsname\relax\def\natexlab#1{#1}\fi
\expandafter\ifx\csname bibnamefont\endcsname\relax
  \def\bibnamefont#1{#1}\fi
\expandafter\ifx\csname bibfnamefont\endcsname\relax
  \def\bibfnamefont#1{#1}\fi
\expandafter\ifx\csname citenamefont\endcsname\relax
  \def\citenamefont#1{#1}\fi
\expandafter\ifx\csname url\endcsname\relax
  \def\url#1{\texttt{#1}}\fi
\expandafter\ifx\csname urlprefix\endcsname\relax\def\urlprefix{URL }\fi
\providecommand{\bibinfo}[2]{#2}
\providecommand{\eprint}[2][]{\url{#2}}

\bibitem[{\citenamefont{Folk et~al.}(2001)\citenamefont{Folk, Holovatch, and
  Yavors'kii}}]{folk01b}
\bibinfo{author}{\bibfnamefont{R.}~\bibnamefont{Folk}},
  \bibinfo{author}{\bibfnamefont{Y.}~\bibnamefont{Holovatch}},
  \bibnamefont{and}
  \bibinfo{author}{\bibfnamefont{T.}~\bibnamefont{Yavors'kii}},
  \bibinfo{journal}{Uspiekhi Fizichieskikh Nauk} \textbf{\bibinfo{volume}{173}},
  \bibinfo{pages}{175} (\bibinfo{year}{2003}) 
  [\bibinfo{note}{cond-mat/0106468}].

\bibitem[{\citenamefont{Harris}(1974)}]{harris74}
\bibinfo{author}{\bibfnamefont{A.~B.} \bibnamefont{Harris}},
  \bibinfo{journal}{J. Phys. C} \textbf{\bibinfo{volume}{7}},
  \bibinfo{pages}{1671} (\bibinfo{year}{1974}).

\bibitem[{\citenamefont{Chayes et~al.}(1986)\citenamefont{Chayes, Chayes,
  Fisher, and Spencer}}]{chayes86}
\bibinfo{author}{\bibfnamefont{J.~T.} \bibnamefont{Chayes}},
  \bibinfo{author}{\bibfnamefont{L.}~\bibnamefont{Chayes}},
  \bibinfo{author}{\bibfnamefont{D.~S.} \bibnamefont{Fisher}},
  \bibnamefont{and} \bibinfo{author}{\bibfnamefont{T.}~\bibnamefont{Spencer}},
  \bibinfo{journal}{Phys. Rev. Lett.} \textbf{\bibinfo{volume}{57}},
  \bibinfo{pages}{2999} (\bibinfo{year}{1986}).

\bibitem[{\citenamefont{Imry and Wortis}(1979)}]{imry79}
\bibinfo{author}{\bibfnamefont{Y.}~\bibnamefont{Imry}} \bibnamefont{and}
  \bibinfo{author}{\bibfnamefont{M.}~\bibnamefont{Wortis}},
  \bibinfo{journal}{Phys. Rev. B} \textbf{\bibinfo{volume}{19}},
  \bibinfo{pages}{3580} (\bibinfo{year}{1979}).

\bibitem[{\citenamefont{Aizenman and Wehr}(1989)}]{aizenman89}
\bibinfo{author}{\bibfnamefont{M.}~\bibnamefont{Aizenman}} \bibnamefont{and}
  \bibinfo{author}{\bibfnamefont{J.}~\bibnamefont{Wehr}},
  \bibinfo{journal}{Phys. Rev. Lett} \textbf{\bibinfo{volume}{62}},
  \bibinfo{pages}{2503} (\bibinfo{year}{1989}), \bibinfo{note}{erratum {\em
  ibid} {\bf 64}, 1311 (1990)}.

\bibitem[{\citenamefont{Tissier}(2002)}]{tissier02b}
\bibinfo{author}{\bibfnamefont{M.}~\bibnamefont{Tissier}}
  (\bibinfo{year}{2002}), \bibinfo{note}{cond-mat/0203370}.

\bibitem[{\citenamefont{Collins and Petrenko}(1997)}]{collins97}
\bibinfo{author}{\bibfnamefont{M.~F.} \bibnamefont{Collins}} \bibnamefont{and}
  \bibinfo{author}{\bibfnamefont{O.~A.} \bibnamefont{Petrenko}},
  \bibinfo{journal}{Can. J. Phys.} \textbf{\bibinfo{volume}{75}},
  \bibinfo{pages}{605} (\bibinfo{year}{1997}).

\bibitem[{\citenamefont{Kawamura}(1998)}]{kawamura98}
\bibinfo{author}{\bibfnamefont{H.}~\bibnamefont{Kawamura}},
  \bibinfo{journal}{J. Phys. C} \textbf{\bibinfo{volume}{10}},
  \bibinfo{pages}{4707} (\bibinfo{year}{1998}).

\bibitem[{\citenamefont{Pelissetto
  et~al.}(2001{\natexlab{a}})\citenamefont{Pelissetto, Rossi, and
  Vicari}}]{pelissetto01}
\bibinfo{author}{\bibfnamefont{A.}~\bibnamefont{Pelissetto}},
  \bibinfo{author}{\bibfnamefont{P.}~\bibnamefont{Rossi}}, \bibnamefont{and}
  \bibinfo{author}{\bibfnamefont{E.}~\bibnamefont{Vicari}},
  \bibinfo{journal}{Phys. Rev. B} \textbf{\bibinfo{volume}{63}},
  \bibinfo{pages}{140414} (\bibinfo{year}{2001}{\natexlab{a}}).

\bibitem[{\citenamefont{Itakura}(2001)}]{itakura01}
\bibinfo{author}{\bibfnamefont{M.}~\bibnamefont{Itakura}},
  \bibinfo{journal}{J. Phys. Soc. Jpn.} \textbf{\bibinfo{volume}{72}},
  \bibinfo{pages}{74} (\bibinfo{year}{2003}).

\bibitem[{\citenamefont{Tissier et~al.}(2001)\citenamefont{Tissier, Delamotte,
  and Mouhanna}}]{tissier01}
\bibinfo{author}{\bibfnamefont{M.}~\bibnamefont{Tissier}},
  \bibinfo{author}{\bibfnamefont{B.}~\bibnamefont{Delamotte}},
  \bibnamefont{and} \bibinfo{author}{\bibfnamefont{D.}~\bibnamefont{Mouhanna}},
  \bibinfo{note}{to appear in Phys. Rev. B} 
  [\bibinfo{note}{cond-mat/0107183}].

\bibitem[{\citenamefont{Garel and Pfeuty}(1976)}]{garel76}
\bibinfo{author}{\bibfnamefont{T.}~\bibnamefont{Garel}} \bibnamefont{and}
  \bibinfo{author}{\bibfnamefont{P.}~\bibnamefont{Pfeuty}},
  \bibinfo{journal}{J. Phys. C: Solid St. Phys.} \textbf{\bibinfo{volume}{9}},
  \bibinfo{pages}{L245} (\bibinfo{year}{1976}).

\bibitem[{\citenamefont{Yosefin and Domany}(1985)}]{yosefin85}
\bibinfo{author}{\bibfnamefont{M.}~\bibnamefont{Yosefin}} \bibnamefont{and}
  \bibinfo{author}{\bibfnamefont{E.}~\bibnamefont{Domany}},
  \bibinfo{journal}{Phys. Rev. B} \textbf{\bibinfo{volume}{32}},
  \bibinfo{pages}{1778} (\bibinfo{year}{1985}).

\bibitem[{\citenamefont{Pelissetto
  et~al.}(2001{\natexlab{b}})\citenamefont{Pelissetto, Rossi, and
  Vicari}}]{pelissetto01b}
\bibinfo{author}{\bibfnamefont{A.}~\bibnamefont{Pelissetto}},
  \bibinfo{author}{\bibfnamefont{P.}~\bibnamefont{Rossi}}, \bibnamefont{and}
  \bibinfo{author}{\bibfnamefont{E.}~\bibnamefont{Vicari}},
  \bibinfo{journal}{Nucl. Phys. B [FS]} \textbf{\bibinfo{volume}{607}},
  \bibinfo{pages}{605} (\bibinfo{year}{2001}{\natexlab{b}}).

\bibitem[{\citenamefont{Br{\'e}zin and Dominicis}(2001)}]{brezin01}
\bibinfo{author}{\bibfnamefont{E.}~\bibnamefont{Br{\'e}zin}} \bibnamefont{and}
  \bibinfo{author}{\bibfnamefont{C.~D.} \bibnamefont{Dominicis}},
  \bibinfo{journal}{Eur. Phys. J. B} \textbf{\bibinfo{volume}{19}},
  \bibinfo{pages}{467} (\bibinfo{year}{2001}).

\bibitem[{\citenamefont{Gross and Sompolinsky}(1985)}]{gross85}
\bibinfo{author}{\bibfnamefont{D.~J.} \bibnamefont{Gross}} \bibnamefont{and}
  \bibinfo{author}{\bibfnamefont{I.~K.~H.} \bibnamefont{Sompolinsky}},
  \bibinfo{journal}{Phys. Rev. Lett.} \textbf{\bibinfo{volume}{55}},
  \bibinfo{pages}{304} (\bibinfo{year}{1985}).

\end{thebibliography}
\end{document}